\documentclass[aps,superscriptaddress,showpacs,floatfix,amsmath,amssymb,twocolumn]{revtex4}
\usepackage{epsfig}
\usepackage{dcolumn}
\usepackage{bm}
\usepackage{graphicx}
\usepackage{colortbl}
\usepackage{rotating}
\usepackage{pdflscape}

\begin{document}

\title{Spectroscopy of $^{26}$F  to probe proton-neutron forces close to the drip line}

\author{A.~Lepailleur}
\affiliation{Grand Acc\'el\'erateur National d'Ions Lourds (GANIL),
CEA/DSM - CNRS/IN2P3, B.\ P.\ 55027, F-14076 Caen Cedex 5, France}
\author{O.~Sorlin}
\affiliation{Grand Acc\'el\'erateur National d'Ions Lourds (GANIL),
CEA/DSM - CNRS/IN2P3, B.\ P.\ 55027, F-14076 Caen Cedex 5, France}
\author{L.~Caceres}
\affiliation{Grand Acc\'el\'erateur National d'Ions Lourds (GANIL),
CEA/DSM - CNRS/IN2P3, B.\ P.\ 55027, F-14076 Caen Cedex 5, France}

\author{B.~ Bastin}
\affiliation{Grand Acc\'el\'erateur National d'Ions Lourds (GANIL),
CEA/DSM - CNRS/IN2P3, B.\ P.\ 55027, F-14076 Caen Cedex 5, France}

\author{C.\ Borcea}
\affiliation{IFIN-HH, P. O. Box MG-6, 76900 Bucharest-Magurele, Romania}
\author{R.\ Borcea}
\affiliation{IFIN-HH, P. O. Box MG-6, 76900 Bucharest-Magurele, Romania}

\author{B.~A.~Brown}
\affiliation{National Superconducting Cyclotron Laboratory and Department of Physics and Astronomy, Michigan
  State University, East Lansing, MI 48824, USA}

\author{ L.\ Gaudefroy}
\affiliation{CEA, DAM, DIF, F-91297 Arpajon, France}

\author{ S.\ Gr\'evy}
\affiliation{Centre d`\'Etudes Nucl\'eaires de Bordeaux Gradignan-UMR 5797, CNRS/IN2P3, Universit\'e de Bordeaux 1, Chemin du Solarium, BP 120, 33175 Gradignan, France}

\author{G. F. ~ Grinyer}
\affiliation{Grand Acc\'el\'erateur National d'Ions Lourds (GANIL),
CEA/DSM - CNRS/IN2P3, B.\ P.\ 55027, F-14076 Caen Cedex 5, France}

\author{G.~Hagen}
\affiliation{Physics Division, Oak Ridge National Laboratory,
Oak Ridge, TN 37831, USA}
\affiliation{Department of Physics and Astronomy, University of
Tennessee, Knoxville, TN 37996, USA}

\author{M.~Hjorth-Jensen}
\affiliation{National Superconducting Cyclotron Laboratory and Department of Physics and Astronomy, Michigan
  State University, East Lansing, MI 48824, USA}
\affiliation{Department of Physics and Center of Mathematics for Applications, University of Oslo, N-0316 Oslo, Norway}

\author{G.~R.~Jansen}
\affiliation{Physics Division, Oak Ridge National Laboratory,
Oak Ridge, TN 37831, USA}
\affiliation{Department of Physics and Astronomy, University of
Tennessee, Knoxville, TN 37996, USA}

\author{O. ~ Llidoo}
\affiliation{Grand Acc\'el\'erateur National d'Ions Lourds (GANIL),
CEA/DSM - CNRS/IN2P3, B.\ P.\ 55027, F-14076 Caen Cedex 5, France}

\author{F.\ Negoita}
\affiliation {IFIN-HH, P. O. Box MG-6, 76900
Bucharest-Magurele, Romania}

\author{F.~ de Oliveira}
\affiliation{Grand Acc\'el\'erateur National d'Ions Lourds (GANIL),
CEA/DSM - CNRS/IN2P3, B.\ P.\ 55027, F-14076 Caen Cedex 5, France}

\author{M.-G.~Porquet}
\affiliation{CSNSM, CNRS/IN2P3 - Universit\'e Paris-Sud, F-91405 Orsay, France}

\author{F.~ Rotaru}
\affiliation {IFIN-HH, P. O. Box MG-6, 76900
Bucharest-Magurele, Romania}

\author{M.-G.~Saint-Laurent}
\affiliation{Grand Acc\'el\'erateur National d'Ions Lourds (GANIL),
CEA/DSM - CNRS/IN2P3, B.\ P.\ 55027, F-14076 Caen Cedex 5, France}

\author{D.\ Sohler}
\affiliation{Institute of Nuclear Research of the
Hungarian Academy of Sciences, P.O. Box 51, Debrecen, H-4001, Hungary}

\author{M.\ Stanoiu}
\affiliation {IFIN-HH, P. O. Box MG-6, 76900
Bucharest-Magurele, Romania} 

\author{J.C.~Thomas}
\affiliation{Grand Acc\'el\'erateur National d'Ions Lourds (GANIL),
CEA/DSM - CNRS/IN2P3, B.\ P.\ 55027, F-14076 Caen Cedex 5, France}

\begin{abstract}
A long-lived $J^{\pi}=4_1^+$ isomer, $T_{1/2}=2.2(1)$ms, has been discovered
at 643.4(1)~keV in the weakly-bound $^{26}_{\;9}$F nucleus. It was populated at GANIL in the
fragmentation of a $^{36}$S beam. It decays by an  internal
transition to the $J^{\pi}=1_1^+$ ground state (82(14)\%),  by $\beta$-decay to
$^{26}$Ne, or beta-delayed neutron emission to $^{25}$Ne. From the
beta-decay studies of the $J^{\pi}=1_1^+$ and $J^{\pi}=4_1^+$ states, new excited
states have been discovered in $^{25,26}$Ne. Gathering the measured
binding energies of the $J^{\pi}=1_1^+-4_1^+$ multiplet in $^{26}_{\;9}$F, we find
that the proton-neutron $\pi 0d_{5/2} \nu 0d_{3/2}$ effective force  used
in shell-model calculations should be reduced  to properly account for
the weak binding of $^{26}_{\;9}$F. Microscopic
coupled cluster theory calculations using interactions derived from chiral effective
field theory are in very good agreement with the energy of the low-lying
$1_1^+,2_1^+,4_1^+$ states in $^{26}$F.  Including three-body forces and coupling to the continuum effects improve the agreement between experiment and theory as compared to the use of two-body forces only. 

\end{abstract}

\maketitle

\emph{Introduction.- } Understanding the boundaries of the nuclear landscape and the origin of magic nuclei  throughout the chart of nuclides  are overarching aims and intellectual challenges in nuclear physics research \cite{Erler}. These are major motivations that drive the developments of present and planned rare-isotope facilities. Studying the evolution of binding energies for the ground and first few excited states in atomic nuclei from the valley of stability to the drip line (where the next isotope is unbound with respect to the previous one) 
is essential to achieve these endeavours. Understanding these trends and providing reliable predictions for nuclei that cannot be accessed experimentally require a detailed understanding of the properties of the nuclear force \cite{sorlin2008,michael2012}.

In the oxygen isotopes, recent experiments have shown that the drip line occurs at the doubly magic  $^{24}$O$_{16}$ \cite{Hoffman09,kanungo2009,Tshoo2012}, as $^{25,26}$O are unbound
~\cite{Hoffman08,lundeberg2012}.  The role of tensor and three-body forces was 
emphasized in ~\cite{Otsuka2010, Otsuka2005} to account for the emergence of the $N=16$ gap at $^{24}$O$_{16}$ and the 'early' appearance of the drip line in the O isotopic chain, respectively. On the other hand, with the exception of  $^{28}$F \cite{christian2012} and $^{30}$F which are unbound, six more neutrons can be added in the F isotopic chain before reaching the drip line at  $^{31}$F$_{22}$~\cite{lukyanov}.  One can therefore speculate that the extension of the drip line between the oxygen and fluorine, as well as the odd-even binding of the fluorine isotopes, arise from a delicate 
balance between the two-body proton-neutron and neutron-neutron  interactions, the coupling to the continuum \cite{Doba07} effects and the three body forces \cite{Hagen2012b,Hagen2012}.\\
\indent The study of $^{26}$F, which is bound by only 0.80(12)~MeV \cite{Jura}, offers a unique opportunity to investigate several aspects of the nuclear force. The $^{26}$F nucleus can be modeled using a simplified single-particle ($s.p.$) description as a closed $^{24}$O core plus a \emph{deeply bound} proton in the $\pi0d_{5/2}$ orbital ($S_{\pi}$($^{25}$F)$\simeq$ -15.1(3)~MeV \cite{Audi}) plus an
\emph{unbound} neutron ($S_{\nu}$($^{25}$O)$\simeq$ 770$^{+20}_{-10}$~keV
\cite{Hoffman08}) in the $\nu0d_{3/2}$ orbital. This simplified picture arises from the fact that the first excited state in
$^{24}$O lies at 4.47~MeV \cite{Hoffman09,Tshoo2012} and the $\pi 0d_{5/2}$
and $\nu 0d_{3/2}$ single particle energies are well separated  from the other orbitals. The low-lying  $J^{\pi}=1_1^+, 2_1^+, 3_1^+,4_1^+$ states in $^{26}$F thus arise, to a first approximation, from the 
interactions of nucleons in the $\pi 0d_{5/2} $ and $ \nu 0d_{3/2}$ orbits.  

Present experimental knowledge concerning the members of the $J^{\pi}=1_1^+, 2_1^+, 3_1^+,4_1^+$
multiplet in $^{26}$F is as follows. A $J^{\pi}=1_1^+$ assignment has been
proposed in \cite{Reed} for the ground state of $^{26}$F from the observation
that its beta decay proceeds to the $J^{\pi}=0_1^+$, $J^{\pi}=2_1^+$ states and a
tentative $J^{\pi}=0_2^+$ state in $^{26}$Ne. The half-life of $^{26}$F was
found to be 10.2$\pm$1.4 ms with a $P_n$ value of 11$\pm$4\%
\cite{Reed}.  A mass excess $\Delta M$ of 18.680(80) MeV was determined for $^{26}$F
in \cite{Jura} using the time-of-flight technique. The $J^{\pi}=2^+_1$ state was discovered 
at 657(7)~keV \cite{Stanoiu12} from the fragmentation of
$^{27,28}$Na nuclei. In addition a charge-exchange reaction with a
$^{26}$Ne beam was used in \cite{Frank} to study unbound states in
$^{26}$F. In this reaction, a neutron capture to the $\nu d_{3/2}$ orbital and a
proton removal from the $\pi d_{5/2}$ (which are both valence orbitals) are
likely to occur leading to the $J^{\pi}=1_1^+-4_1^+$ states. 
The resonance  observed at 271(37) keV above the neutron emission threshold \cite{Frank}  could tentatively be attributed to the $J^{\pi}=3_1^+$ in $^{26}$F, as it was the only state of the $J^{\pi}=1_1^+-4_1^+$ which was predicted to be unbound. With the determination of the binding energies of the $J^{\pi}=1^+_1-3_1^+$ states, the only missing information is the energy of the $J^{\pi}=4_1^+$ state.  In this Letter, we demonstrate that the  4$_1^+$ state is  isomeric and decays by competing internal transition and 
$\beta$ decay. Its binding energy is determined and those of the $1^+_1 - 2^+_1$ states are re-evaluated.
The comparison of the measured binding energies of the  $J^{\pi}=1_1^+-4_1^+$ states 
with two theoretical approaches, the nuclear shell model and Coupled Cluster (CC) theory, provides a stringent test of the nuclear forces, where a large proton-to-neutron binding energy asymmetry is present.

\emph{Experiment.- } The $^{26}$F nuclei were produced through the fragmentation of a 77.6
MeV/A $^{36}$S$^{16+}$ primary beam with a mean intensity of 2 $\mu$Ae
in a 237 mg/cm$^2$ Be target. They were selected by the LISE \cite{LISE}
spectrometer at GANIL, in which a wedge-shaped degrader of 1066 $\mu$m
was inserted at the intermediate focal plane. The produced nuclei
were identified from their energy loss in a stack of Si
detectors and by their time-of-flight with respect to the GANIL
cyclotron radio frequency. The production rate of $^{26}$F was 
6~pps with a purity of 22\% and a momentum acceptance of 2\%. Other transmitted nuclei, ranked by
decreasing order of production, were $^{28}$Ne, $^{29}$Na, $^{27}$Ne,
$^{24}$O, $^{22}$N and $^{30}$Na. They were implanted in a 1~mm-thick
double-sided Si stripped detector (DSSSD) composed of 256 pixels (16 strips in
the X and Y directions) of 3$\times$3~mm$^2$-each located at the
final focal point of LISE. This detector was used to detect the
$\beta$-particles in strips $i,i\pm$1 following the implantation of a
radioactive nucleus in a given pixel $i$.
With an energy threshold of $\sim$80~keV in the individual strips, a
$\beta$-efficiency of 64(2)\% was achieved for  $^{26}$F which
was implanted at central depth of the DSSSD.  The $\beta$-efficiency has been determined from the comparison of the intensity of a given $\gamma$-ray belonging to the decay of $^{26}$F gated or not on a $\beta$-ray. Four clover Ge
detectors of the EXOGAM  array \cite{EXOGAM} surrounded the DSSSD to detect
the $\gamma$-rays, leading to a $\gamma$-ray efficiency of 6.5\% at 1~MeV.

\begin{figure}
\centering \epsfig{width=8.5cm,height=7cm,file=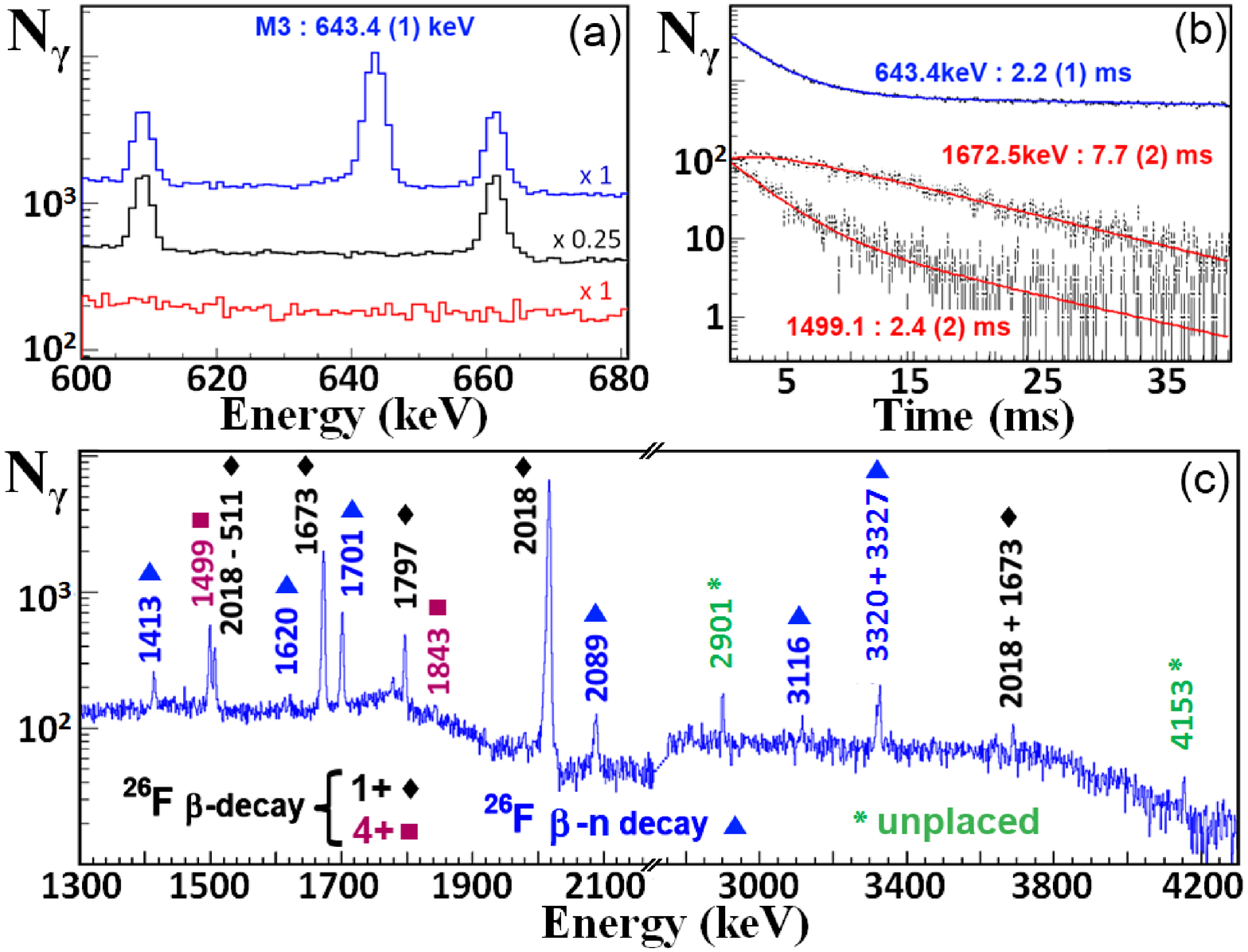} 
\caption{(Color online) \textbf{(a)}:  $\gamma$-ray spectra obtained up to 2 ms after the implantation of $^{26}$F  (upper spectrum), or after the implantation of any nucleus except $^{26}$F (middle spectrum). The bottom spectrum shows the $\beta$-gated $\gamma$-rays following the implantation of $^{26}$F. \textbf{(b)} Time spectra 
between implanted $^{26}$F and $\gamma$-rays, from which half-lives were deduced.  The 643.4(1) keV and $4^+ \rightarrow 2^+_1$ (1499.1(4) keV) transitions have the same half-life, while the one gated on the $2^+_2 \rightarrow 2^+_1$ (1672.5(3) keV) transition has a larger half-life. \textbf{(c)}:  $\beta$-gated $\gamma$-ray spectrum following the implantation of $^{26}$F  up to 30 ms. Symbols  and colors indicate which lines correspond to the $\beta$-decay of the 1$^+$ ($\blacklozenge$,black) and 4$^+$ ($\blacksquare$, red) or to the  $\beta$ delayed-neutron branch ($\blacktriangle$, blue). The same color codes are used in the decay scheme of Fig. \ref{26Fdecay}.  Two lines ($\ast$, green) could not be placed in the decay scheme  of $^{26}$F.}
\label{26Fisom}
\end{figure}

The $\gamma$-ray spectra obtained up to 2~ms after the implantation of
a radioactive nucleus are shown in Fig.~\ref{26Fisom}(a). In this frame the
upper (middle) spectrum is obtained by requiring that $^{26}$F (all
except $^{26}$F) precedes the detection of a $\gamma$ ray. A delayed
$\gamma$-ray transition at 643.4(1)~keV is clearly observed after the
implantation of $^{26}$F. The bottom spectrum of Fig.~\ref{26Fisom} (a)
is operated in similar condition than
the top one, with the additional requirement that $\gamma$-rays are
detected in coincidence with a $\beta$ transition. As the 643.4(1)~keV
is not in coincidence with $\beta$ particles  it must correspond to an internal transition
($IT$) de-exciting an isomeric state in $^{26}$F, which has a half-life of
2.2(1) ms (see Fig.~\ref{26Fisom}(b)). This isomer is likely the $4^+$
state we are searching for. It either decays directly to the $1^+$
ground state, hereby establishing the $4^+$ state at 643.4(1)~keV.
Alternatively, the 643.4(1)~keV energy may correspond, but with a weak
level of confidence, to the 657(7)~keV state observed
in~\cite{Stanoiu12}. In this hypothesis, the isomerism of the $4^+$
state would be due to the emission of a very low energy $4^+ \rightarrow
2^+$ transition (up to 10 keV to ensure having a long-lived isomer), then
followed by the $2^+ \rightarrow 1^+$ transition. In either case, the
excitation energy of the $4^+$ state lies at approximately 650(10) keV.

\begin{figure}
\centering \epsfig{width=8.5cm,height=8.5cm,file=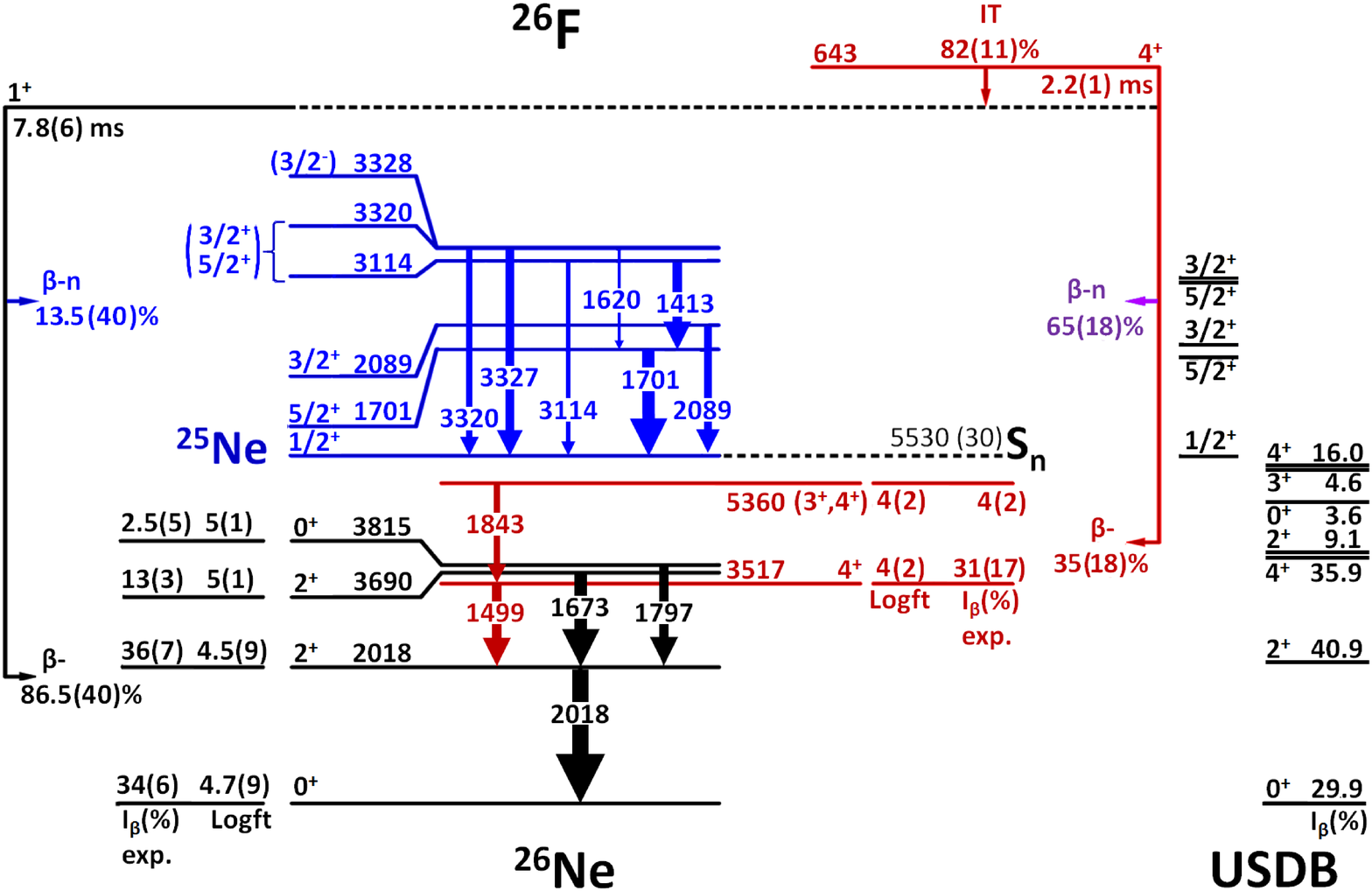} 
\caption{(Color on line) Decay scheme obtained from the decays of the $4^+$ (red) and $1^+$ states (black) in $^{26}$F to  $^{26}$Ne and $^{25}$Ne (blue). Shell model predictions obtained with the USDB 
interaction are shown in the right hand side.}
\label{26Fdecay}
\end{figure}

The decay of this $4^+$ state occurs through a competition between
an internal transition ($IT$) and $\beta$-decay to two states in
$^{26}$Ne. The half-lives corresponding to the $IT$ (2.2(1) ms) as well as to the 1499.1(4)~keV (2.4(2)ms) and 1843.4(8)~keV (2(1)ms) peaks of Fig.~\ref{26Fisom}(c) are the same.  These two transitions are seen in mutual coincidences, as
well as with the 2017.6(3)~keV $\gamma$-ray, previously assigned to the
$2^+_1 \rightarrow 0^+_1$ transition in $^{26}$Ne in \cite{Reed}. This
establishes two levels at 3516.7(4)~keV and 5360.1(9)~keV in $^{26}$Ne as shown
in Fig. \ref{26Fdecay}. Following the Gamow-Teller $\beta$-decay
selection rules the  $4^+$ isomer should mainly proceed
to the $J^{\pi}=4_1^+$ state in the vibrator nucleus $^{26}$, which we attribute to  
the 3516.7(4) keV state.

All other observed transitions in Fig. 1(c) from $^{26}$F belong to the decay of the $1^+$ ground state, as their half-lives differ significantly from that of
the $4^+$ isomeric state. The two $\gamma$-ray transitions at 1672.5(3) keV and 1797.1(4) keV
were found to be in coincidence with the 2017.6(3) keV transition, but not in
mutual coincidence. This establishes two levels at 3690.1(4) keV and 3814.7(5)
keV which have compatible half-lives of 7.7(2)~ms, and 7.8(5) ms,
respectively.  These states presumably belong to the two-phonon multiplet of
states $J^{\pi}=0^+_2, 2^+_2, 4^+_1$ among which the 3516.7(4)~keV one was
assigned to $J^{\pi}=4^+_1$ (see above). Using in-beam $\gamma$-ray spectroscopy
from the fragmentation of a $^{36}$S beam \cite{Bell}, the
feeding of the 3516.7(4)~keV level was the largest, that of the 3689.8(4) keV
state was weaker, while the state at 3814.7(5) keV was not fed.  As this method
mainly produces Yrast states, i.e. states having the highest spin
value in a given excitation energy range, we ascribe  $J^{\pi}=2^+_2$ 
to the state at 3690.1(4) keV, in accordance with \cite{Gibe07}, and  $J^{\pi}=0^+_2$  to the state at 3814.7(5) keV. The fitting of the
decay half-lives must include the direct $1^+_1$ decay of $^{26}$F
as well as the partial feeding from the $4_1^+ \rightarrow 1_1^+$ transitions.
This leads to a growth at the beginning of the time spectrum
(Fig.~\ref{26Fisom} (b) for the 1673 keV $\gamma$-ray) which depends on the isomeric
ratio $R$ and on the internal transition coefficient $IT$. These
parameters are furthermore constrained by the amount of the 643.4(1) keV
$\gamma$-rays observed per implanted $^{26}$F nucleus, leading to $R$= 42(8)\% and
$IT$=82(11)\%.

The $\beta$ feedings derived from the observed $\gamma$-ray intensities are given in 
Fig. \ref{26Fdecay}. In the $\beta$-delayed neutron branch of $^{26}$F to $^{25}$Ne, some levels observed in \cite{Reed,Padgett,Catford} are confirmed, while a new state is proposed at 3114.1(8) keV as the 1413.2(7) keV and 1700.9(4) keV $\gamma$-rays are in coincidence and the summed $\gamma$-ray energy is observed at 3116(2) keV. A $P_n$ value of 16(4)\% (consistent with $P_n$=11(4)\% \cite{Reed}) is extracted for $^{26}$F from the observation of the 979.7 keV $\gamma$-ray  in the grand-daughter nucleus $^{25}$Na whose branching ratio of 18.1(19)\% was determined in \cite{Goosman}.  We therefore adopt a mean value of $P_n$=13.50(40)\% for $^{26}$F. The proposed level scheme and branching ratios agree relatively well with the shell-model calculation shown on the right side of Fig. \ref{26Fdecay}.

\begin{figure}
\centering \epsfig{width=8.5cm,file=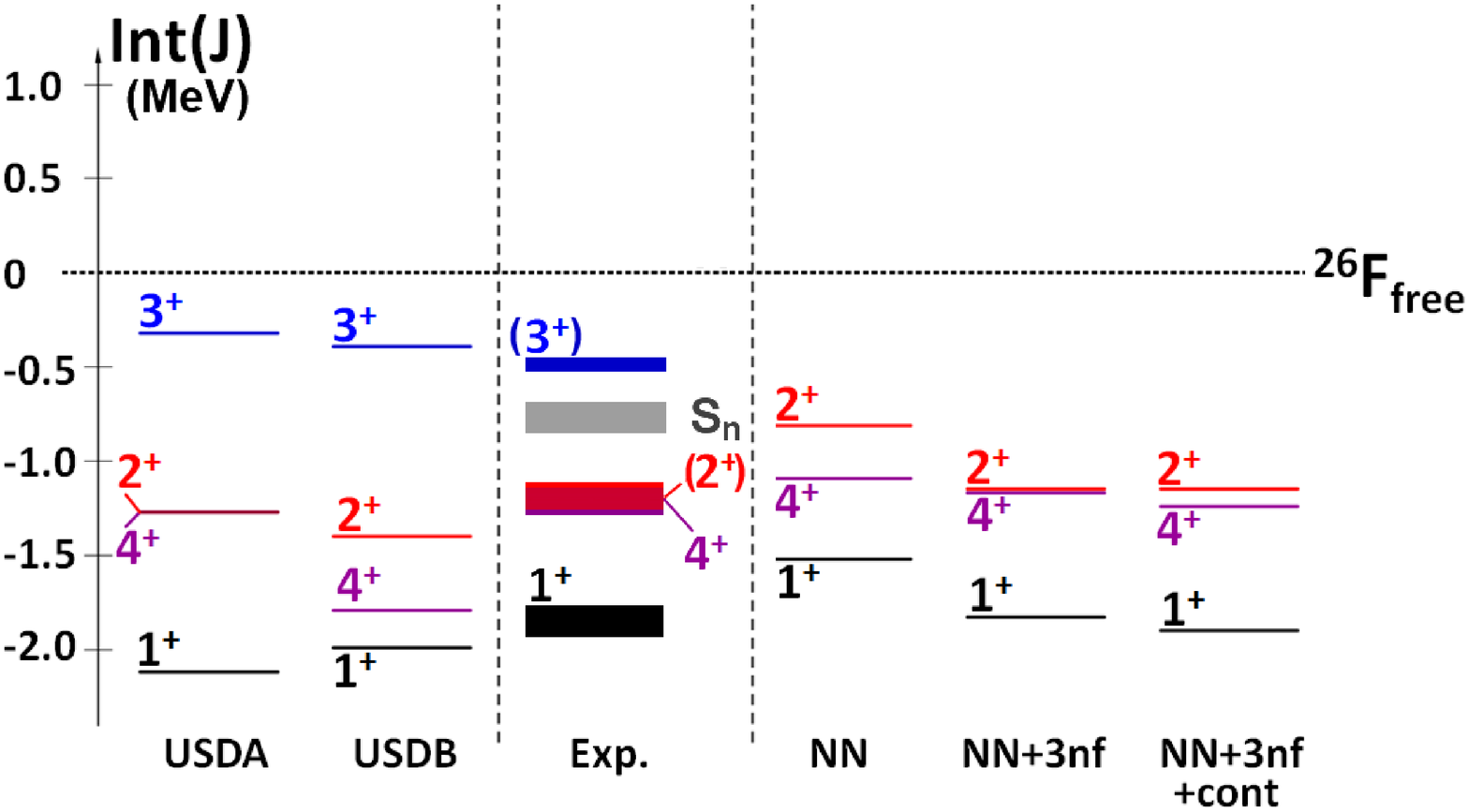}
\caption{(Color online) Calculated and experimental interaction energies $\mathrm{Int(1-4)}$ in MeV  in $^{26}$F. Shell-model calculations are shown in the first column  using the USDA or USDB interactions, while the third column shows results obtained with CC calculations. Experimental results are in the center. The thickness of the lines corresponds to $\pm 1 \sigma$ error bar.} \label{intJ}
\end{figure}

The discovery of this new isomer has an important consequence on the
determination of the atomic mass of the $^{26}$F ground state as well on the interpretation of the one-neutron knock-out cross sections from $^{26}$F of Ref. \cite{Rodriguez}.  It is very likely that the measured atomic
mass of Ref. \cite{Jura} corresponds to a mixture of the ground  and the isomeric states (unknown at that time).  As the $^{26}$F nuclei were produced in the
present work and that of \cite{Jura} in similar fragmentation
reactions involving a large number of removed nucleons, we can
reasonably assume that the $^{26}$F isomeric ratio is the same in the
two experiments.  The shift in the $^{26}$F atomic mass as a function of the
isomeric ratio $R$ amounts to -6.43 keV/\%, which for $R$=42(8)\% yields -270(50) keV.

\emph{Discussion.- } The comparison between the experimental binding energies of these states can now be made with two theoretical approaches, the nuclear shell model and  CC theory. 
The experimental (calculated) interactions elements arising from the coupling between a $d_{5/2}$ proton  and a $d_{3/2}$ neutron, 
labeled $\mathrm{Int(J)}$,  are  extracted from the experimental (calculated) binding energies BE as
\[
\mathrm{Int(J)}= \mathrm{BE(^{26}F)_J} - \mathrm{BE(^{26}F_{free})}. 
\]
In this expression BE($^{26}$F$_{free}$) corresponds to the binding energy of the $^{24}$O+1p+1n system, in which the valence proton in the $d_{5/2}$ orbit and the neutron in the $d_{3/2}$ orbit do not interact. It can be written as
\[
\mathrm{BE(^{26}F_{free})} = \mathrm{BE(^{25}F)}_{5/2^+}+\mathrm{BE(^{25}O)}_{3/2^+}-\mathrm{BE(^{24}O)}_{0^+}.
\]

Using the relative binding energy of +0.77$^{+20}_{-10}$ MeV \cite{Hoffman08} between
$^{24}$O and $^{25}$O , the measured atomic masses in $^{25}$F and
$^{26}$F~\cite{Jura}, and the shift in energy due to the isomeric content (see above)
 it is found that the experimental value of $\mathrm{Int(1)}$ is -1.85(13)~MeV. The values of $\mathrm{Int(2)}$= -1.19(14) MeV and $\mathrm{Int(4)}$=-1.21(13) MeV 
are obtained
using the $J^{\pi}=2_1^+$ and $J^{\pi}=4_1^+$ energies of 657(7)keV and 643.4(1) keV, respectively. 
A value of $\mathrm{Int(3)}$=-0.49(4) MeV is derived  from the energy of the $J^{\pi}=3_1^+$ resonance with respect to the
$^{25}$F ground state.

In the shell-model calculations of Refs. ~\cite{usd,usdab}, the two-body matrix elements corresponding to interactions in the $sd$ valence space are fitted to reproduce properties of known nuclei. Applying these interactions to nuclei not included in the global fits (such as bound and unbound states in $^{26}$F) implies that shell-model calculations towards the drip lines can be viewed as predictions.
Due to the strong 
coupling to the continuum, and a likely absence of  many-body correlations not included in the fits, these interactions may fail in reproducing properties of nuclei like $^{26}$F.
Owing to its simple structure, $^{26}$F provides a unique possibility
to probe the strength of the proton-neutron interaction close to the drip line. The wave functions of the $J^{\pi}=1_1^+-4_1^+$ states are composed of mainly ($80-90\%$) pure  $\pi 0d_{5/2} \otimes \nu 0d_{3/2}$
component. By calculating all states in the $J^{\pi}=1_1^+-4_1^+$ multiplet, it can be seen in Fig. \ref{intJ} that the $J^{\pi}=1_1^+$ state is less bound than calculated 
by about 17\% (8\%) and that the multiplet of experimental states is
compressed by about 25\% (15\%) compared with the USDA (USDB) calculations. This points to a weakening of the residual interactions, which caused the energy splitting between the members of the multiplet.

We have also performed microscopic CC
\cite{Coester1958,Coester1960} calculations for $^{26}$F. This method
is particularly suited for nuclei with closed (sub-)shells, and their
nearest neighbors. Moreover, CC theory can easily handle nuclei in which protons 
and neutrons have significantly different binding energies. 
To estimate the $\pi 0d_{5/2}-\nu 0d_{3/2}$ interaction
energy ($\mathrm{Int(J)}$), we use CC theory with singles and doubles
excitations with perturbative triples corrections
\cite{Kucharski1998,Taube2008} for the closed-shell nucleus
${}^{24}$O, the particle-attached CC method for ${}^{25}$O and ${}^{25}$F \cite{Hagen2010} and the 
two-particle attached formalism for $^{26}$F \cite{Jansen2011}.  
We employ interactions from chiral effective field theory \cite{Entem2003}. The
effects of three-nucleon forces are included as corrections to
the nucleon-nucleon interaction by integrating one nucleon in the
leading-order chiral three-nucleon force over the Fermi sphere with a
Fermi momentum $k_{\mathrm{F}}$ in symmetric nuclear matter \cite{Holt2009}. The
parameters recently established in the
oxygen chain \cite{Hagen2012} are adopted in the present work. 
We use a Hartree-Fock basis built from $N_{\mathrm{max}} = 17$ major spherical
oscillator shells with the oscillator frequency $\hbar\omega = 24$~MeV. This is sufficiently large to achieve convergence of the calculations
 for all isotopes considered.  Using two-body nucleon-nucleon forces we get the ground-state energy of $^{26}$F at $-173.2$ MeV which is underbound by $\sim~ 11$ MeV compared to
experiment. However, the relative spectra for the excited states are
in fair agreement with experiment (see Fig.~\ref{intJ}). 
In order to account for the coupling to the continuum in $^{26}$F, we use a
real Woods-Saxon basis  for the $\nu 1s_{1/2}$ and $\nu 0d_{3/2}$ partial waves~\cite{Jensen2011}. The inclusion
of continuum effects and three-nucleon forces  improve the situation, the ground state energy is at -177.07 MeV, and the low-lying spectra is in very good agreement with experiment.  
The $J^\pi = 3^+$ state in $^{26}$F is a resonance and  to compute this state we need a 
Gamow-Hartree-Fock basis~\cite{Michel2009}. We are currently working on generalizing the two-particle attached
CC implementation to a complex basis. Therefore, the interaction energy of the $J=3$ state is not shown in Fig.~\ref{intJ}. Consistently with the shell-model calculations described above, a simple picture emerges from the microscopic CC calculations: about 85\% of the $1^+-4^+$ wave functions are composed of $1s0d$-shell components, in which configurations consisting of the  $\pi 0d_{5/2}$ and $\nu 0d_{3/2}$ s.p. states play a major role. 

\emph{Conclusions.- } To summarize, a new $J^{\pi}=4_1^+$ isomer with a 2.2(1) ms half-life has been discovered at 643.4(1) keV. Its isomeric decay to the $J^{\pi}=1_1^+$ ground state and $\beta$-decay to the
$J^{\pi}=4_1^+$ state in $^{26}$Ne were observed. Gathering the $\beta$-decay
branches observed from the $J^{\pi}=1_1^+$ and $J^{\pi}=4_1^+$ states, partial level
schemes of $^{26}$Ne and $^{25}$Ne were obtained.  In addition, the $^{26}$F nucleus is a benchmark case for studying proton-neutron interactions far from stability.  The experimental states $J=1^+ - 4^+$ arising from the $\pi d_{5/2} \otimes \nu d_{3/2}$ coupling in $^{26}_{9}$F are more compressed  than the USDA and USDB shell model results. The experimental $J^{\pi}=1_1^+,2_1^+,4_1^+$ states are less bound as well. These two effects point to a dependence of the effective two-body interaction used in the shell model as a function of the proton-to-neutron binding energy asymmetry. Coupled-cluster calculations including three-body forces and coupling to the particle continuum are in excellent agreement with experiment for the bound low-lying states in $^{26}$F. 

\acknowledgments {\small 
This work was partly supported by the Office of Nuclear Physics, U.S. Department
of Energy (Oak Ridge National Laboratory), the NSF grant PHY-1068217, the  OTKA contract K100835, the grant of the Romanian National Authority for Scientific Research, CNCS Ð UEFISCDI, PN-II-RU-TE-2011-3-0051 as well as 
by FUSTIPEN (French-U.S. Theory Institute for Physics with Exotic Nuclei) under DOE grant number DE-FG02-10ER41700. This research used computational
resources of the National Center for Computational Sciences,
the National Institute for Computational Sciences, and the Notur project in Norway. 
The CENBG  is greatly acknowledged for the loan of the DSSSD detector. A. L. thanks V. Tripathi for communicating information which we used for calibration purposes.}


\begin{thebibliography}{0}
\expandafter\ifx\csname natexlab\endcsname\relax\def\natexlab#1{#1}\fi
\expandafter\ifx\csname bibnamefont\endcsname\relax
  \def\bibnamefont#1{#1}\fi
\expandafter\ifx\csname bibfnamefont\endcsname\relax
  \def\bibfnamefont#1{#1}\fi
\expandafter\ifx\csname citenamefont\endcsname\relax
  \def\citenamefont#1{#1}\fi
\expandafter\ifx\csname url\endcsname\relax
  \def\url#1{\texttt{#1}}\fi
\expandafter\ifx\csname urlprefix\endcsname\relax\def\urlprefix{URL }\fi
\providecommand{\bibinfo}[2]{#2}
\providecommand{\eprint}[2][]{\url{#2}}

\end{thebibliography}


\begin{thebibliography}{100}


\bibitem{Erler} J. Erler {\em et al.}, Nature \textbf{486}, 509 (2012).
\bibitem{sorlin2008} O. Sorlin and M.-G. Porquet, Prog. Part. Nucl. Phys. \textbf{61},  602 (2008), 
Phys. Scr. (2012) in press.
\bibitem{michael2012} T.~Baumann, A.~Spyrou, and M.~Thoennessen, Rep. Prog. Phys. {\bf 75},  036301  (2012).

\bibitem{Hoffman09} C.~R.~Hoffmann {\em et al.}, Phys.~Lett.~B {\bf 672}, 17 (2009).
\bibitem{kanungo2009} R.~Kanungo {\em et al.}, Phys.~Rev.~Lett.~{\bf 102}, 152501 (2009).
\bibitem{Tshoo2012} K. Tshoo  {\em et al.}, Phys.~Rev.~Lett.~{\bf 109}, 022501 (2012).  
\bibitem{Hoffman08} C. R. Hoffmann {\em et al.}, Phys. Rev. Lett. {\bf 100}, 152502 (2008).
\bibitem{lundeberg2012} E.~Lunderberg {\it et al.}, Phys.~Rev.~Lett.~{\bf 108}, 142503 (2012). 

\bibitem{Otsuka2010} T. Otsuka {\it et al.}, Phys.~Rev.~Lett.~{\bf 105}, 032501 (2010). 
\bibitem{Otsuka2005} T. Otsuka {\it et al.}, Phys. Rev. Lett. {\bf 95} 232502 (2005).
\bibitem{christian2012} G.~Christian {\em et al.}, Phys.~Rev.~Lett. {\bf 108}, 032501 (2012).
\bibitem{lukyanov}  S. M. Lukyanov and Yu. E. Penionzhkevich, Physics of Atomic Nuclei
{\bf 67}, 1627 (2004) and references herein.

\bibitem{Doba07} J. Dobaczewski et al., Prog. Part. Nucl. Phys. 59 (2007) 432
\bibitem{Hagen2012b} G.~Hagen, M.~Hjorth-Jensen, G.~R.~Jansen, R.~Machleidt, and T.~Papenbrock, Phys.~Rev.~Lett. {\bf 109}, 032502 (2012).
\bibitem{Hagen2012} G.~Hagen, M.~Hjorth-Jensen, G.~R.~Jansen, R.~Machleidt, and T.~Papenbrock, Phys. Rev. Lett. {\bf 108}, 242501 (2012).
\bibitem{Jura} B.~Jurado {\em et al.}   Phys.~Lett.~B. {\bf 649}, 43  (2007).
\bibitem{Audi} G. Audi  {\em et al.}, Nucl. Phys. A {\bf 729}, 3  (2003).
\bibitem{Reed} A.~T.~Reed {\em et al.}, Phys.~Rev.~C {\bf 60}, 024311 (1999).
\bibitem{Stanoiu12} M. Stanoiu {\em et al.}, Phys.~Rev.~C {\bf 85}, 017303 (2012).
\bibitem{Frank} N.~Frank {\em et al.}, Phys.~Rev.~C {\bf 84}, 037302 (2011). 
\bibitem{LISE} R. Anne and A.C. Mueller, Nucl. Inst. Meth. B {\bf 70} 276 (1999)
\bibitem{EXOGAM} J. Simpson et al., Acta Phys. Hung., New Series, Heavy Ion Physics {\bf 11} 159 (2000).
\bibitem{Bell} M. Belleguic {\em et al.}, Phys.~Rev.~C {\bf 72}, 054316 (2005). 
\bibitem{Gibe07} J. Gibelin {\em et al.}, Phys.~Rev.~C {\bf 75}, 057306 (2007).
\bibitem{Padgett} S.~W.~Padgett {\em et al.}, Phys.~Rev.~C {\bf 72}, 064330 (2005).
\bibitem{Catford} W.~Catford {\em et al.}, Phys.~Rev.~Lett. {\bf 104},  192501 (2010).
\bibitem{Goosman} D. R. Goosman {\em et al.}, Phys. Rev. {\bf C 7}, 1133 (1973).
\bibitem{Rodriguez} C. Rodr\'iguez-Tajes  {\em et al.}, Phys. Rev. {\bf C 82}, 024305 (2010).
\bibitem{usd} B.~A.~Brown and B.~H.~Wildenthal, Ann.~Rev.~Nucl.~Part.~Sci.~{\bf 38},  29 (1988).
\bibitem{usdab} B.~A.~Brown and W.~A.~Richter, Phys.~Rev.~C {\bf 74},  034315 (2006).
\bibitem{Coester1958} F.~Coester, Nucl. Phys. {\bf 7}, 421 (1958).
\bibitem{Coester1960} F.~Coester and H. K\"ummel, Nucl. Phys. {\bf 17}, 477 (1960).
\bibitem{Kucharski1998} S.~A.~Kucharski and  R.~J.~Bartlett, J. Chem. Phys. {\bf 108}, 5243 (1998).
\bibitem{Taube2008} A.~G.~Taube and R.~J.~Bartlett, J. Chem. Phys, {\bf 128}, 044110 (2008).
\bibitem{Hagen2010} G.~Hagen {\em et al.}, Phys. Rev. C {\bf 82}, 034330 (2010).
\bibitem{Jansen2011} G.~R.~Jansen{\em et al.},  Phys. Rev. C {\bf 83}, 054306 (2011).
\bibitem{Entem2003} D.~R.~Entem and R.~Machleidt, Phys. Rev. C {\bf 68}, 041001 (2003).
\bibitem{Holt2009} J.~W.~Holt, N.~Kaiser, and W.~Weise, Phys. Rev. C {\bf 79}, 054331 (2009); Phys. Rev. C {\bf 81}, 024002 (2010).
\bibitem{Jensen2011}  \O{} Jensen {\em et al.}, Phys.~Rev.~Lett. {\bf 107},  032501 (2011).
\bibitem{Michel2009} N.~Michel {\em et al.},  J.~Phys.~G {\bf 36}, 013101 (2009).

\end{thebibliography}
\end{document}